# Plasmon-Enhanced Photoresponse of a Single Silver Nanowire and its Networked Devices

Mohammadali Razeghi[1], Merve Üstünçelik[1], Farzan Shabani[1], Hilmi Volkan Demir[1,2,3,4], T. Serkan Kasırga[1,2*]

[1] Institute of Materials Science and Nanotechnology – UNAM, Bilkent University, Ankara 06800, Turkey

[2] Department of Physics, Bilkent University, Ankara 06800, Turkey

[3] Department of Electrical and Electronics Engineering, Bilkent University, Ankara 06800, Turkey

[4] LUMINOUS! Centre of Excellence for Semiconductor Lighting and Displays, The Photonics Institute, School of Electrical and Electronic Engineering, School of Physical and Mathematical Sciences, School of Materials Science and Engineering Nanyang Technological University, Singapore 639798, Singapore

*Corresponding author email: kasirga@unam.bilkent.edu.tr

**Abstract**

Photo-bolometric effect is critically important in optoelectronic structures and devices employing metallic electrodes with nanoscale features due to heating caused by the plasmonic field enhancement. One peculiar case is individual silver nanowires (Ag NWs) and their networks. Ag NW-networks exhibit excellent thermal, electrical, and mechanical properties, providing a simple yet reliable alternative to common flexible transparent electrode materials used in optoelectronic devices. To date there have been no reports on the photoresponse of Ag NWs. In this work, we show that a single Ag NW and a network of such Ag NWs possess a significant, intrinsic photoresponse thanks to the photo-bolometric effect, as directly observed and measured using scanning photocurrent microscopy. Surface plasmon polaritons (SPP) created at the contact metals or plasmons created at the nanowire-metal structures cause heating at the junctions where a plasmonic field enhancement is possible. The local heating of the Ag NWs results in negative photoconductance due to the bolometric effect. Here an open-circuit response due to the plasmon-enhanced Seebeck effect was recorded at the NW-metal contact junctions. The SPP-assisted bolometric effect is found to be further enhanced by decorating the Ag NWs with Ag nanoparticles. These observations are relevant to the use of metallic nanowires in plasmonic applications in particular and in optoelectronics in general. Our findings may pave the path for plasmonics enabled sensing without a spectroscopic detection.

**Main Text**

Silver nanowire (Ag NW) networks offer excellent flexibility, transparency, and electrical conductivity in optoelectronics[1]. Moreover, excellent plasmon coupling and conducting properties of Ag NWs make them appealing candidates for plasmonics[2–5]. The creation and control of surface plasmon polariton (SPP) modes in Ag NWs allows light concentration at the nanoscale. Local electric field intensity at a nanowire junction can be greatly enhanced owing to the plasmonic effects[6] and this enhancement of field localization can raise the local temperature of junctions by several tens to hundreds of Kelvins depending on the incident light power[7–9]. Indeed, the local temperature at such hot spots is sufficient to weld Ag NWs to each other to obtain the desired network properties for the optoelectronic applications[10–12].

The photoresponse from a material can be generated by the separation of non-equilibrium charge carriers and/or by the photo-thermal effects[13–18]. Typically, a photoresponse is unexpected in metals as the non-equilibrium carriers rapidly thermalize. The photo-thermal effects such as the photo-bolometric effect and/or the Seebeck effect produce insignificant photoresponse due to the efficient heat transport mechanisms. In the photo-bolometric effect, the light induced temperature change must be large enough to induce an electrical resistivity change of the metal. Similarly, in the Seebeck effect the junction between two metals with dissimilar Seebeck coefficients should be hotter than the opposing ends of the metals, so that an emf (electromotive force) can be generated. In both cases, efficient heat transport pathways in metals results in small temperature alterations that hinders a measurable photo-thermal response. Yet, at the nanoscale, various mechanisms might be at play to result in an optical response in metals[17,19,20]. As an instance, weak photoconductance is observed in an alkali-ion-intercalated two-dimensional gold nanoparticle (NP) array because of the SPP-assisted bolometric enhancement of the conductivity of these NPs in the array[21]. Similarly, Au NPs on a $TiO_2$ photoanode have shown plasmonic heating assisted photoresponse during water splitting[22].

Here, employing scanning photocurrent microscopy (SPCM), we systematically investigate the photoresponse of individual Ag NWs and their networks. We discovered that when a Ag NW network is under bias, a considerable negative photoresponse to the light with the responsivity of up to ~130 mA/W is generated. This is less than an order of magnitude lower than that of silicon PN junctions[23]. We investigated the origin of the photoresponse measured both in single Ag NW devices and networks. Our studies conclusively show that the photoresponse originates from the photo-bolometric effect, enhanced by the plasmon-induced heating at the nanowire junctions. To maximize the plasmonic field-enhancement effect, we further decorated Ag NWs and NW networks with Ag NPs and showed that the measured photo-bolometric effect can be enhanced substantially by up to two orders of magnitude. We also observed a weak open-circuit response from the single NWs as well as their networks and explained the underlying mechanisms using finite element method simulations.

SPCM is a powerful technique in determining the photoresponse of nanowires and nanotubes[19,20]. In SPCM, spatial maps of photo-induced electrical current and the corresponding reflected light intensity are created by raster-scanning a diffraction-limited laser spot focused over the sample. Electrical probes collect the photogenerated current and the silicon detector located at the reflection path of the laser beam reads the reflected intensity. This enables spatially-resolved identification of the photoresponse from the device under investigation. A schematic of this SPCM setup and the images created by the setup are presented in Figure 1a. Further details of our experimental setup are given in the Supporting Information.

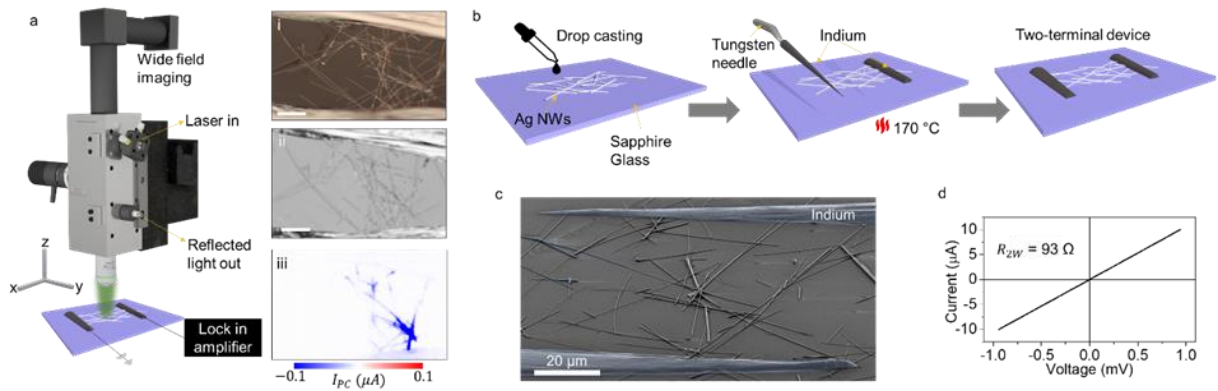

**Figure 1 a.** Schematic of our SPCM setup: The continuous-wave laser beam focused to a diffraction-limited spot raster-scans the device. (i) The wide field reflection image, (ii) reflected light intensity map and (iii) generated photocurrent map of a typical device given in the panels on the right. The scale bars are 20 μm. **b.** The device fabrication steps depicted: Ag NWs in IPA solution is drop-casted on a sapphire or glass substrate. The substrate with Ag NWs is heated to 170 °C and indium needles are placed over the nanowire network using a micromanipulator. Upon cooling, the device is ready for two-terminal electrical and scanning photocurrent measurements. **c.** False-color SEM image of a typical device: Indium needles are shaded with blue. **d.** Current vs. voltage graph shows the 2-wire resistance, $R_{2W}$, of a typical device.

First, we measured the photoresponse of the Ag NW network devices. Polyol process synthesized Ag NWs in isopropyl alcohol (IPA) solution (Sigma Aldrich, 0.5 wt%) were used. Ag NWs were characterized using X-ray diffraction (XRD), Raman spectroscopy, and X-ray photoelectron spectroscopy (XPS). Raman spectrum collected from the nanowires reveals that polyvinylpyrrolidone (PVP) molecules from the synthesis covers the Ag NW surface (**Figure S2a**)[24]. PVP plays a part in the formation of the nanowires and protects Ag NWs from the ambient[24]. XRD pattern (**Figure S2b**) matches with the face-centered-cubic Ag (Ref. Card 98-005-3759) and finally XPS shows the metallic nature of the Ag nanowires (**Figure S2c**) (details are in the Supporting Information).

To fabricate the network devices, 0.02 wt. % Ag NW suspension was drop-casted onto a sapphire substrate (**Figure 1b**). In the case of fabrication of single nanowire devices, the suspension was diluted four times. Once the IPA evaporated, indium needles were placed as the electrical contacts. To place the indium needles on the Ag NW network, the substrate was heated to above 160 °C for several minutes (**Figure 1b**). Once the indium needle contacted the hot substrate, it melted and formed the electrical connection with the Ag NW network. The high-temperature processing also significantly reduces the contact resistance as compared to the gold electrode devices before annealing[25]. Despite the heating step, some devices exhibited large resistances on the order of MΩs. After biasing the devices, most devices reduce to few tens of Ωs. **Figure 1c** displays the false-colored SEM micrograph of a typical device and **Figure 1d** shows its IV curve with the two-wire resistance of $R_{2W} = 93$ Ω. Our Ag NW networks remained stable for an extended duration of up to 2 months after the fabrication.

SPCM maps taken on a Ag NW network device (Device #1) under 0, 100, and -100 mV are depicted in **Figures 2a**, **b** and **c**, respectively (at excitation wavelength $\lambda = 532$ nm and with laser power, $P = 100$ μW). The sign of the photocurrent ($I_{PC}$) indicates the direction of the photo-induced current with respect to the lock-in amplifier. When the signal was into (out of) the lock-in amplifier, we registered a positive (negative) photocurrent. The reflection map corresponding to the SPCM maps is shown in **Figure 2d**. 0 mV measurement shows that there is a weak photocurrent generation at certain parts of the network. We speculate that this is because of the plasmon-assisted heating of the Ag NW junctions that leads to the Seebeck effect. A more detailed analysis is provided in the following paragraphs.

When a bias was applied across the sample, a negative photoconductance ($G_{Ph} = \frac{I_{PC}}{V}$, where $V$ is the applied bias) was observed along the NWs as well as in the vicinity of and at the NW-NW and NW-contact junctions. The local extrema of the photocurrent reside at the NW-NW junctions. Intriguingly, there is a considerable response from the indium contacts. The negative photoconductance over the indium contacts increases towards the nanowire junctions. It is unplausible for optical absorption mechanisms to increase temperature increase in the thick (~1 μm) indium contacts and result in a measurable photo-bolometric response from the indium needles. For the similar reasons, without an enhanced heating mechanism the Seebeck effect from the indium-Ag NW junction is unexpected. Moreover, despite Ag NWs exhibit a high temperature coefficient of resistance of ~2 K$^{-1}$ above 50 K [26], it is not possible for the laser powers employed in our experiments to lead to a measurable photo-bolometric response without a mechanism leading to local hot spot creation.

To explain the origins of the photoresponse observed in Ag NW networks, we conducted a series of controlled experiments. First, it is worth noting that the photoconductance is the same for different biases at points where there is no open circuit photoresponse, as shown in **Figure 2e**. This photoconductance corresponds to an overall increase in the resistance of the network, $\delta R \approx -I_{PC}\frac{R_{2W}^2}{V}$. For Device #1, $\delta R = 0.052\ \Omega$. This value is consistent with the finite element method simulation results (**Figure S3a-b**). The bolometric response is simulated with a point heat source located at the NW-NW junction. The local temperature rise of 5 K may result in a bolometric response of 50 mΩ, which is a plausible temperature rise for the laser powers employed in our experiments. We also measured $-G_{Ph}$ vs. the incident laser power, given in **Figure 2f**. The linear relation hints at a photo-thermal origin for the photoresponse[17,19,27,28].

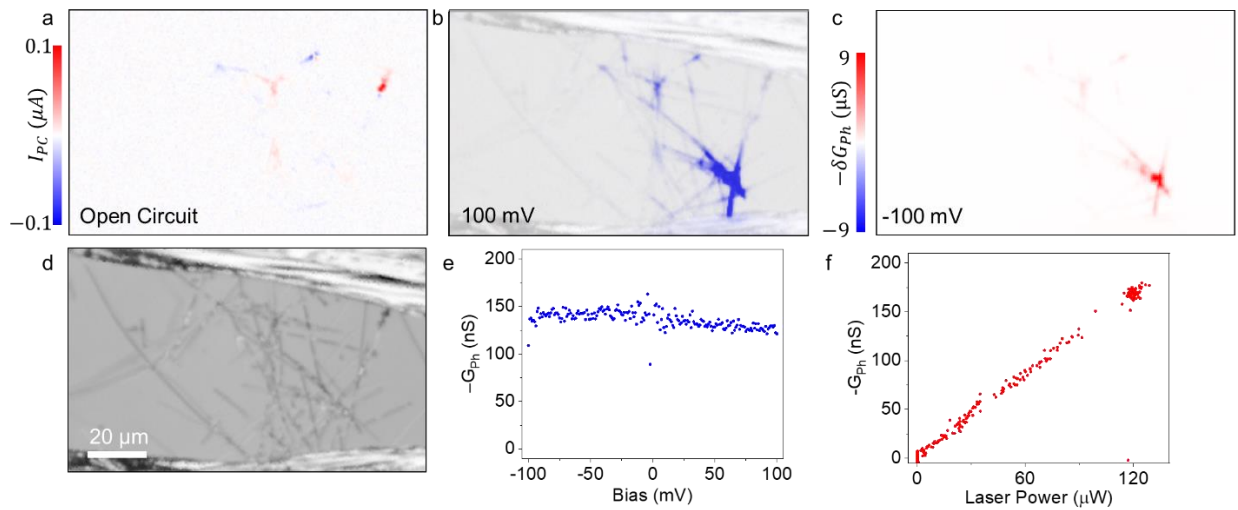

**Figure 2** SPCM maps of a network device under **a.** open-circuit (0mV), **b.** 100 mV (overlayed over the reflection map for better positioning of the photoresponse locations). The scale is saturated to make all the features apparent. The maximum $-I_{PC}$ is 0.9 µA. It is noteworthy that there is a significant photoconductance change over the indium contacts as well as nanowires throughout the entire network. **c.** $-\delta G_{Ph}$ map plotted under -100 mV bias. **d.** Reflection map shows the details of the network between the indium contacts. **a, b, c** and **d** have the same scale bar. **e.** Negative photoconductance vs. applied bias measurement taken from a different network device shows a very small variation in the photoconductance across different biases. **f.** Negative photoconductance increases linearly with the increasing laser power. The gap in the data points between 90 µW and 120 µW is due to gap in the variable neutral density filter.

Next, we fabricated single nanowire devices and measured their photoresponse for a better understanding of the photoconductance in the network devices. We fabricated devices with two different contact configurations. In the first device configuration, the nanowires were placed over pre-patterned gold electrodes (**Figure 3a**). **Figure 3b** shows the reflection map (**Figure 3b-I**) and the corresponding $-G_{Ph}$ (**Figure 3b-II**). There is a small open-circuit photocurrent ($I_{PC}^0$) over the contact regions marked on **Figure 3b-II**. **Figure 3c-I** displays the bias dependence of $-I_{PC}$ in which $-I_{PC}$ linearly increases with the bias. $-G_{Ph}$ exhibits a logarithmic growth (**Figure 3c-II**). When the open-circuit response is subtracted from the rest of the photocurrent values, the photoconductance, $-\frac{I_{PC}-I_{PC}^0}{V}$ remains constant for the different biases. This is similar to what we observe in the network devices.

In the second device configuration, indium needles were placed over a single nanowire (as depicted in **Figure 1b**). **Figure 3d** presents a false-color SEM image of a device. **Figure 3e** shows the reflection map (**Figure 3e-I**) and the corresponding photoconductance map (**Figure 3e-II**) of the indium contacted single nanowire device (SEM images of the device are provided in the supporting information, **Figure S4**). In both device configurations strong negative photoconductance at the metal contact-nanowire junction region, mostly over the electrodes is observed. This hints that the photoconductance modulation comes from the plasmonic heating effects at the naturally occurring gaps between the nanowires and the contact metals as other photoresponse mechanisms are unplausible.

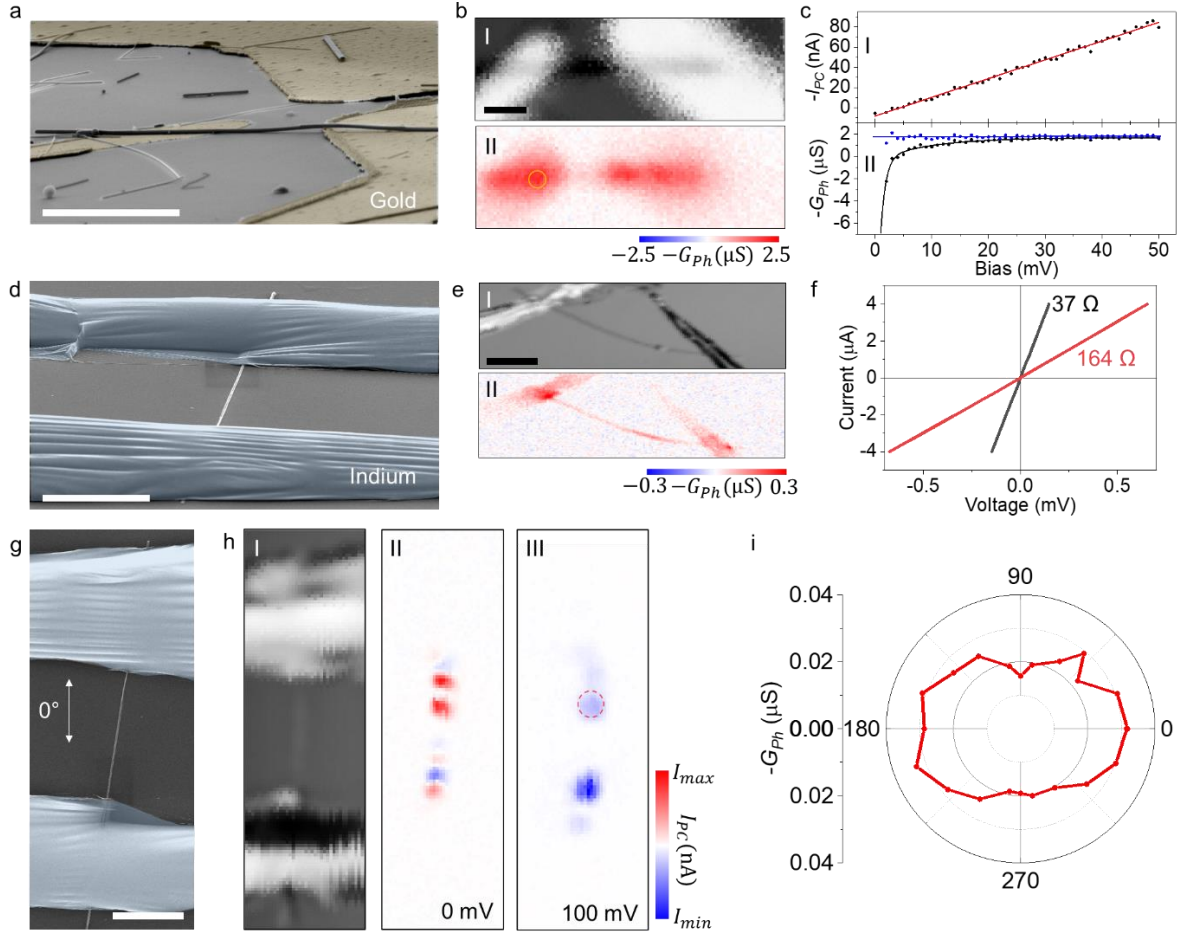

**Figure 3 a.** False colored SEM image of an Ag NW on pre-patterned gold electrodes. Scale bar is 5 µm. **b.** Reflection map (I) and the corresponding $-G_{Ph}$ map under 20 mV bias (II) showing the photoconductance over the device. It is noteworthy that there is a significant conductance modulation over the gold electrodes. A 532 nm laser source was used at 65 µW power. The yellow circle indicates the position where $-I_{PC}$ vs. bias measurements were taken from. The scale bar is 5 µm. **c.** $-I_{PC}$ vs. bias (I) graph shows a linearly increasing $-I_{PC}$ with the bias (black dots). The red line is the linear fit to the data. Zero bias response is ~6 nA. Lower panel (II) shows $-G_{Ph}$ (black dots) and $-\frac{I_{PC}-I_{PC}^0}{V}$ (blue dots) vs. bias. Solid lines are fits to the data. **d.** False colored SEM image of indium contacted Ag NW. Note the grating-like structures over the indium that naturally occur due to thermal contraction upon cooling. Scale bar is 5 µm. **e.** Reflection map (I) and the corresponding $-G_{Ph}$ map under 100 mV bias (II) showing the conductance modulation over the indium contacts. Notice that the modulation intensifies near the indium-NW junction. Also, the visible defect in the reflection map results in enhanced plasmonic response which is clear in the photoconductance map. Data was collected with a 532 nm laser at 96 µW. Scale bar is 10 µm. **f.** Current vs. voltage graphs of the same devices, before (red dots) and after (black dots) reduction of the photoconductance. **g.** False colored SEM image of a device with the incident polarization angle marked. **h.** (I) Reflection map, (II) open-circuit response, and (III) 100 mV response maps are shown. $I_{max} = 0.25$ nA for (II), 25 nA for (III) and $I_{min} = -0.25$ nA for (II) and $-25$ nA for (III). **i.** The polar chart shows the incident polarization angle vs. $-G_{Ph}$ at 100 mV bias. The data was collected from the individual maps taken around the red dashed circle in **h(III)**.

Now we would like to hypothesize and elucidate the mechanisms leading to the photocurrent generation over the electrode region near the electrode-NW junction. To explain photocurrent generation mechanism over the metal electrodes, we hypothesize that the SPPs created at the rough contact metal surfaces reach the junction between the Ag NW and the contact metal and generate localized hot spots. As briefly discussed earlier, Ag NWs exhibit a high temperature coefficient of resistance and the temperature rise at the junctions can result in a bolometric response. **Figure S5** shows the bolometric response of a network device under homogenous heating. Despite the relatively small laser powers employed in our experiments, the junctions can be heated up by a few to few tens of Kelvins[8,10,29] and this may lead to a measurable change in the overall photoconductance of the nanowire. For the plasmon-enhanced heating, SPP creation is needed. The creation of SPP requires matching the wave vector of the incident photon to that of the SPP modes[30]. This can be achieved by frustrating the incident electromagnetic wave through surface roughness or with grating structures on the metal surface[31,32]. Moreover, SPP creation is highly dependent on the orientation of the incident light polarization.

To test the hypothesis stated in the previous paragraph, we conducted a series of experiments. First, we would like to note that the contact resistance plays a detrimental role in determining the photoresponse of metallic materials[17]. Due to the PVP coverage of Ag NWs, making electrical connection with small contact resistance requires welding of NWs to each other and to the contacts[12]. To begin with, we measured the 2-wire resistance of a single nanowire device as 164 Ω before the SPCM experiments. After 81 SPCM mappings under various testing conditions, the photoresponse disappeared and the 2-wire resistance dropped down to 37 Ω (**Figure 3f**). The significantly improved resistance value shows that the contact resistance of the device is reduced by fusing the contact metal to the Ag NW[33]. It can be speculated that the contact resistance improvement is due to the SPP-assisted heating generated at the contact region or due to Joule heating as the device is under bias during SPCM measurements[34]. Another possibility that may lead to the decrease in the resistance is the interface reconstruction due to electromigration in Ag NWs[35]. Regardless of the possible mechanisms, the naturally occurring gaps between the NW and the contact close and the SPP-assisted bolometric effect emerging at the contacts disappears.

Another observation that supports the SPP-assisted heating at the NW-contact junction hypothesis is that the photoresponse is generated mostly over the gold surface and the indium contacts in single NW devices and in the vicinity of the NW-metal junction. The response from the metal junctions exhibits the same photoconductance sign and their relative intensities can be dramatically different from each other. This indicates that the phenomena that lead to the optical response is independent of the contact configuration of the devices. Among the known photoresponse mechanisms, only the sign of the signal generated by the bolometric effect is not affected by the contact configuration. Also, it is worth noting that grating-like structures that naturally form on the surface of the indium needles, as shown in the SEM image in **Figure 3d**, can provide an efficient path for SPP creation. The period of the features is mostly uniform over the region of interest, and for the device shown in **Figure 3d**, the grating period is ~500 nm.

To further demonstrate the role of the SPP-enhanced bolometric effect at the nanowire-metal junctions, we investigated the polarization dependence of the observed photoresponse. We performed these experiments by rotating the polarization angle of the incident beam with respect to the Ag NW. For the polarization-dependent experiments, we used a 632.8 nm HeNe laser source. Details of the polarization-dependent measurements are provided in the Supporting

Information. **Figure 3g** shows the SEM image of a nanowire we tested for the polarization-dependent photoresponse. SPCM reflection map and the photocurrent maps at 0 and 100 mV biases obtained with 532 nm excitation are given in **Figure 3h**. Under zero bias, $I_{PC}$ alternates between negative and positive responses with a period of 2.4 µm. When 100 mV is applied, $I_{PC}$ becomes -25 nA with no positive signal remaining. **Figure 3i** shows the polar plot of the incident polarization vs. $-G_{Ph}$ taken from the red dashed circle in **Figure 3h-III**. The strong incident polarization angle dependence of the photoresponse suggests plasmonic nature of the observed signal.

Now, we focus on the open circuit-response. Open-circuit response was observed in both the networked and single nanowire devices. As the network devices consist of many NW-NW junctions we focused our attention on simpler single NW devices as seen in **Figure 3h-II**. We consider this response to be due to the Seebeck effect because of the SPP-assisted heating of the Ag-In junctions. Signal along the crystal is likely to be due to the undulations of the silver nanowires after biased measurements. Further explanation to the alternating photoresponse is provided in the supporting information. The Seebeck coefficient difference between Ag and In is $\Delta S_{Ag-In} \approx 0.3$ µV/K [36,37]. The device we measured in Figure 3g-i has 6 Ω resistance. This means the emf generated by the Seebeck effect is ~1 nV. Based on $\Delta S_{Ag-In}$ we can estimate that the junction temperature is raised by ~5 mK for 100 µW laser power in this sample.

The plasmonic coupling to the incident laser beam can be enhanced by decorating the nanowires with nanoparticles[38–40]. This will create more SPPs and plasmonic heating at the nanoparticle-NW gap will be distributed across the length of a nanowire, which should increase the bolometric response observed as compared to the bare Ag NW devices, where the signal only emerges from the contact-NW junction. To test the idea, we fabricated an Ag NW network device on glass substrate and decorated it with Ag nanoparticles (NPs) of 20 ± 5 nm in diameter (see SI for details)[41]. **Figure 4a** shows the saturated SPCM map (with indium contacts outlined) to reveal fainter features that lie outside of the contacts as well as over the indium. **Figure 4b** shows the unsaturated map of the section outlined in **Figure 4a**. The device reported here exhibits 130 mA/W responsivity, determined by the ratio of the generated photocurrent over the incident laser power. This is more than an order of magnitude larger than the response from the networked device reported in **Figure 2**, where responsivity is measured as 9 mA/W.

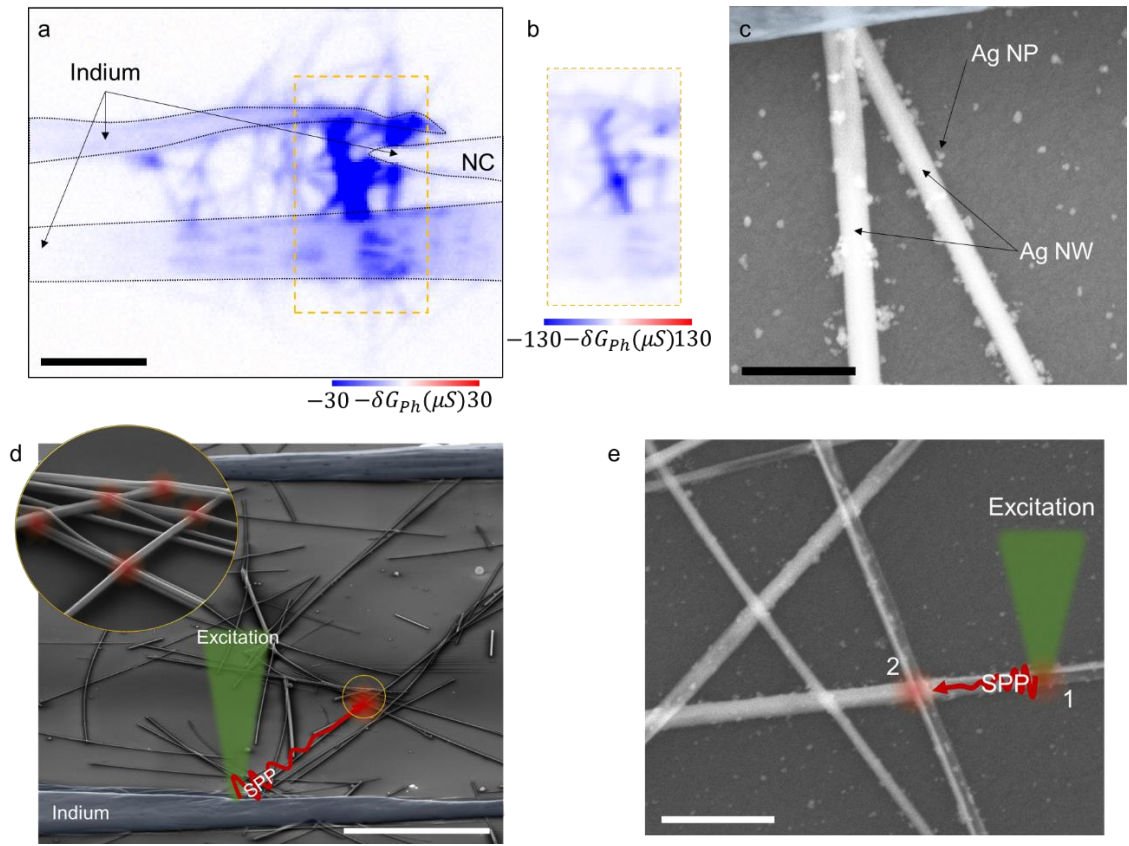

**Figure 4 a.** Photoconductance map of an Ag NP decorated network device under -100 mV bias. Outlines of the indium contacts are marked with black dashed lines. The needle on the right side marked with NC is an indium needle that was not connected to the device electrically. The map scale is saturated to reveal the photoconductance change over the Ag NWs outside of the device and over the indium needles. Maps were taken with a 532 nm laser at 93 µW. The scale bar is 10 µm. **b.** Unsaturated $-\delta G_{Ph}$ map of the area marked with the yellow dashed rectangle in **a**. **c.** Close-up SEM micrograph of Ag NWs decorated with Ag NPs. Indium contact is false-colored with blue. The scale bar is 500 nm. **d.** SPP-assisted bolometric response is depicted on a false-colored SEM image. The excitation creates SPPs at the contacts or regions where the light coupling to the metal structure is possible. The SPPs travel to the NW-NW junctions and the local field enhancement creates hot spots. The scale bar is 20 µm. **e.** A similar mechanism is depicted on the SEM image for Ag NP decorated Ag NWs. The light coupling is possible near the nanowires thanks to the Ag NPs which create two hot spots, marked with 1 and 2 on the SEM image. The scale bar is 1 µm.

Here, as we predicted, all the nanowires connected to the indium exhibit a photoresponse. This is due to the enhanced light coupling to the metallic structures thanks to the Ag NPs in the vicinity of the Ag NWs and shows that the photoresponse is not a local response, rather SPPs traveling to the relevant junctions cause the photoresponse over the nanowires as explained in undecorated devices. Also, the response is spread all over the Ag NWs and the response is significantly enhanced thanks to the heating that takes place due to the field enhancement at the NW-NP junction. **Figure 4c** shows a close-up SEM image of the Ag NP decorated network device in which Ag NPs are clearly visible in the vicinity of the NWs. The other interesting observation consistent with the undecorated devices is the enhanced photoresponse from the grating-like features over the indium contacts. **Figure 4d** depicts the creation of SPPs over the metal contacts and creation of local hot spots at the NW-NW junctions that results in the bolometric response observed in the device. Similarly, **Figure 4e** depicts the coupling of the incident light to the SPPs to create a hot spot both locally and at the NW-NW junction at a distance.

In summary, here we reported the photoresponse from single Ag NWs and their networks. Our findings reveal that the plasmonic effects result in enhanced photo-bolometric response and the Seebeck response in single Ag NWs and their networks. Moreover, decorating the Ag NWs with Ag nanoparticles results in a modified photoresponse. These findings may enable a facile way to fabricate and more importantly measure all-metallic photo-sensitive optoelectronic devices and sensors. As such, attaching capture antibodies over the Ag NW network may result in a detectable change in the photocurrent generation in the presence or in the absence of the target molecules[42].

**Author Contributions**

T.S.K. designed and conceded the experiments. M.R. and M.U. performed the experiments with the help of T.S.K. F.S. and H.V.D. provided the nanoparticles. All the authors discussed the results and contributed to the writing of the manuscript.

**Acknowledgements**

T.S.K. acknowledges funding from Scientific and Technological Research Council of Turkey (TUBITAK) under grant number 118F061. We would like to thank Prof. Hüsnü Emrah Ünalan for providing useful feedback on the manuscript and the Ag NWs for the preliminary experiments. We also would like to thank Prof. Fatih İnci for his useful comments on the manuscript.

# Supporting Information: Plasmon-Enhanced Photoresponse of a Single Silver Nanowire and its Networked Devices

Mohammadali Razeghi[1], Merve Üstünçelik[1], Farzan Shabani[1], Hilmi Volkan Demir[1,2,3,4], T. Serkan Kasırga[1,2*]

[1] Institute of Materials Science and Nanotechnology – UNAM, Bilkent University, Ankara 06800, Turkey

[2] Department of Physics, Bilkent University, Ankara 06800, Turkey

[3] Department of Electrical and Electronics Engineering, Bilkent University, Ankara 06800, Turkey

[4] LUMINOUS! Centre of Excellence for Semiconductor Lighting and Displays, The Photonics Institute, School of Electrical and Electronic Engineering, School of Physical and Mathematical Sciences, School of Materials Science and Engineering Nanyang Technological University, Singapore 639798, Singapore

*Corresponding author email: kasirga@unam.bilkent.edu.tr

**Scanning Photocurrent Microscopy Experimental Setup**

The scanning photocurrent microscope (SPCM) used in the experiments is a commercially available setup from LST Scientific Inst. Ltd. The setup consists of a scanning microscope that simultaneously provides a wide field view and the reflected light intensity. The laser beam is focused with an Olympus 40x 0.6NA ULWD objective with compensation collar. This creates a ~400 nm resolution for 532 nm excitation wavelength, determined by Gaussian fitting the first derivative of the intensity profile across an ultra-steep metal contact edge. The scanner can take 25 nm steps and scan size can be selected from 100 nm to 2.5 cm. The laser power incident on the sample is continuously monitored and registered by a calibrated power meter. The incident laser power can be tuned with a variable neutral density filter. The laser beam is chopped at 2 kHz via a mechanical chopper for phase sensitive detection of the photocurrent via a lock-in amplifier. The generated current is amplified with an SR570 current pre-amplifier, and the output of the amplified signal is fed to SR 830 lock-in amplifier. This allows very high sensitivity of the detected photo-response down to a few picoamperes. The bias is applied through the SR570 or an external voltage source.

The polarization dependent studies are performed using 632.8 nm HeNe laser source unless otherwise stated. We also performed similar measurements with 532 nm laser. The laser beam produced by both HeNe and 532 nm source are linearly polarized. To obtain a high extinction ratio, we aligned a linear polarizer with the polarization of the laser source. To rotate the polarization angle with respect to the sample, we used a quarter wave plate (QWP) with a linear polarizer (LP) on a rotational mount as the analyzer. **Figure S1a** shows the schematic of the polarization control configuration in the SPCM setup. **Figure S1b** shows laser power vs. analyzer angle. The intensity variation is less than 1% except 330°. The polarization angle dependent photocurrent and photoconductance data reported in the main text is read from the SPCM photocurrent maps. This eliminates the possible laser beam shift caused signal changes.

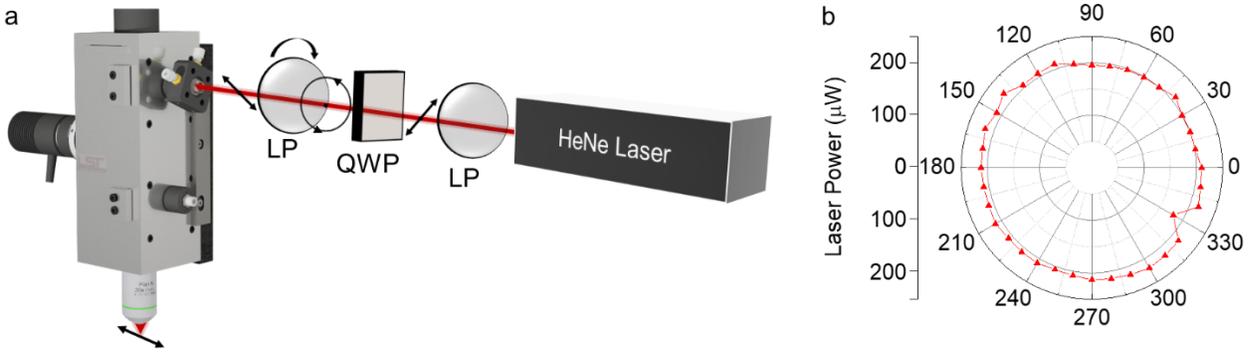

**Figure S1 a.** SPCM configuration with the polarization control optics. **b.** Laser power vs. analyzer angle shows that the laser power change for different angles is very small.

### Characterization of Silver Nanowires

For the experiments reported in this paper, we used commercial Ag NWs from Sigma-Aldrich (Product code: 778095). Nanowires are 120-150 nm in diameter and 20-50 μm in length and provided as 0.5% isopropyl alcohol (IPA) suspension.

We used Raman spectroscopy, XRD and XPS to characterize the nanowires. **Figure S1a** shows the Raman spectrum taken from a dense region of nanowires. Ag NWs exhibit a featureless Raman spectrum, however, the polyvinylpyrrolidone (PVP) coverage over the nanowires due to the inherent residue of the polyol process shows three pronounced peaks at 237, 1337 and 1605 cm$^{-1}$. We also observed a broad peak around 2900 cm$^{-1}$. These peaks agree with those reported in the literature. **Figure S1b** shows the XRD θ-2θ scan of the silver nanowires deposited on glass substrate. Finally, XPS survey given in **Figure S1c** shows that the full width at half maximum for the Lorentzian fits to the Ag $3d_{5/2}$ and Ag $3d_{3/2}$ peaks are ~0.6 eV and no other Lorentzian can be fitted under the peaks. Thus, we infer that the freshly deposited nanowires are in pure metal form.

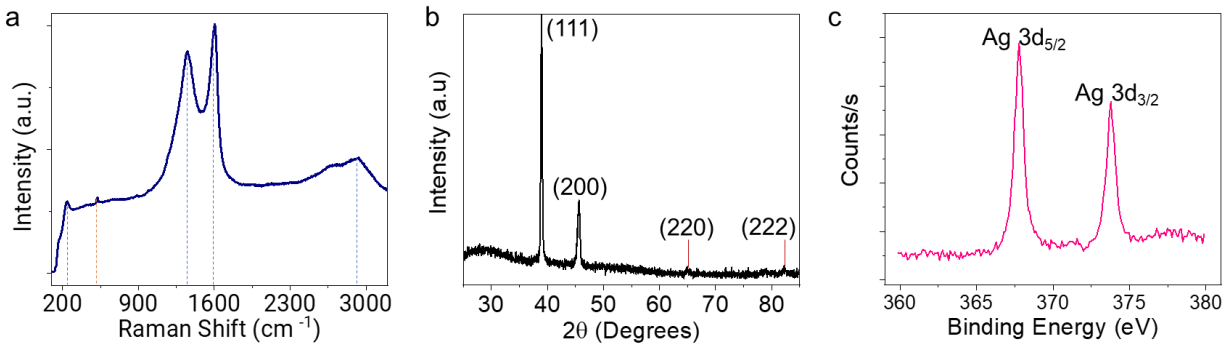

**Figure S2 a.** Raman spectrum taken from an Ag NW. Blue dashed lines indicate the PVP spectra and the orange dashed line shows the Raman mode of the substrate. **b.** XRD scan shows all the major diffraction peaks of Ag NWs. **c.** XPS spectrum of Ag 3d peaks. The peaks can be fitted with single Lorentzian function, showing that there are no oxidation states for freshly prepared devices.

### Finite Element Method Simulations of Resistance Change in Silver Nanowires

We performed finite element method (FEM) simulations to obtain an estimate on how much temperature rise is required to reach the experimentally measured resistance change in the devices. We used COMSOL Multiphysics simulation package to perform these simulations. Two

20 µm long cylindrical nanowires with 150 nm diameter was modelled over a glass slate. The ends of the nanowires were fixed at the ambient temperature. A thermal contact was defined across the glass substrate and the Ag NW. To model the SPP assisted local heating, we defined a point heat source with a 100 nm radius at the junction of the NWs. **Figure S3a** shows the temperature distribution over the nanowire. **Figure S3b** shows the resistance change ($\delta R$) and the maximum temperature rise ($\delta T_{max}$) under varied heat input.

We also would like to provide a theoretical analysis for the bolometric response of the device shown in **Figure 3e**. For a small temperature rise $\delta T(x)$, where coordinate x runs from 0 to 1 from right to left contact, the conductance change can be estimated as $G_{Ph} \approx -\frac{1}{R^2}\frac{dR}{dT}\delta T_l$. Here $\delta T_l$ is the temperature change under the laser spot and $R(T)$ is the resistance of the nanowire. With this formula, we can estimate the order of magnitude of $G_{Ph}$ for the device reported in Figure 3e. If we take the resistance as ~160 Ω and $\frac{dR}{dT} \approx 0.4\ \Omega/K$ based on the Resistance vs. temperature graph and $\delta T_l \approx 0.5\ K$ from the Comsol simulations, we obtain $G_{Ph} \approx$ -0.7 µS which is slightly larger than our measurements. Here, $\delta T_l$ value might be overestimated as the Comsol simulation assumed the heating at the junction of two nanowires.

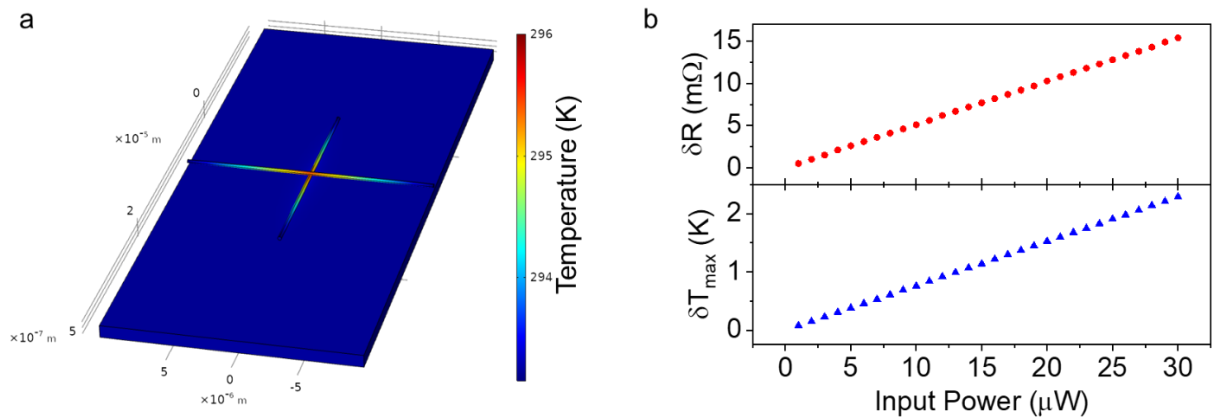

**Figure S3 a.** Thermal surface map of the NW modelled on a glass substrate under 30 µW heat input. **b.** The graph shows the resistance change $\delta R$ in the upper panel and the maximum temperature on the NW, $\delta T_{max}$, in the lower panel. $\delta R$ and $\delta T_{max}$ values agree with the experimentally determined values reported in the main text.

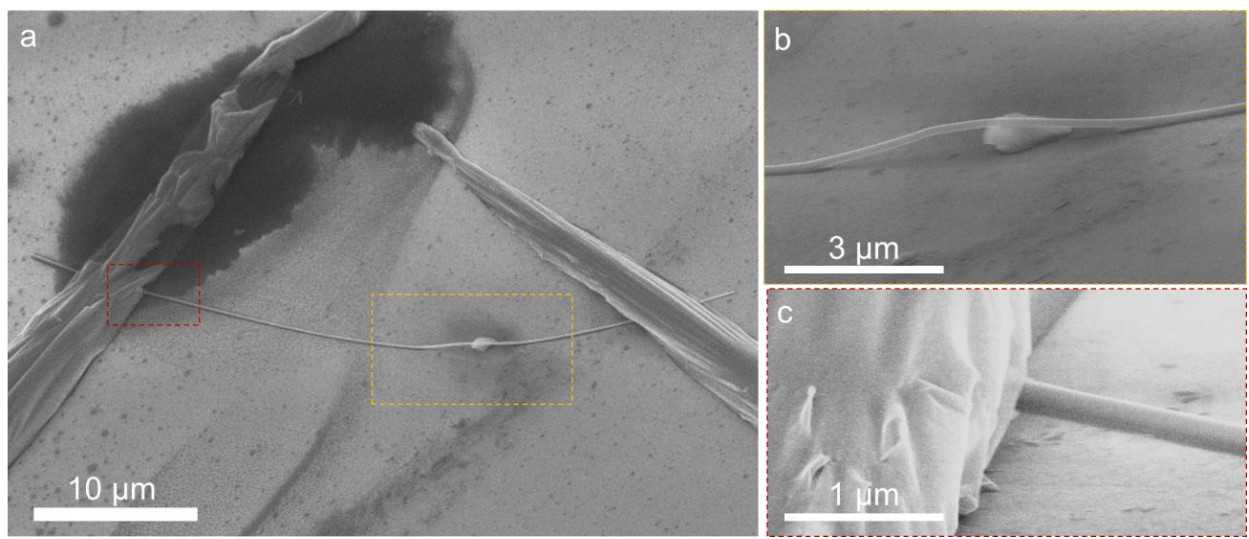

**Figure S4 a.** SEM image of the device measured in Figure 3e in the main text. Colored regions correspond to zoomed in sections shown in **b** and **c**. **b.** Bulged section of the nanowire is shown. The photoresponse is enhanced in the vicinity of the bulged region. **c.** Close-up image of the NW-indium junction.

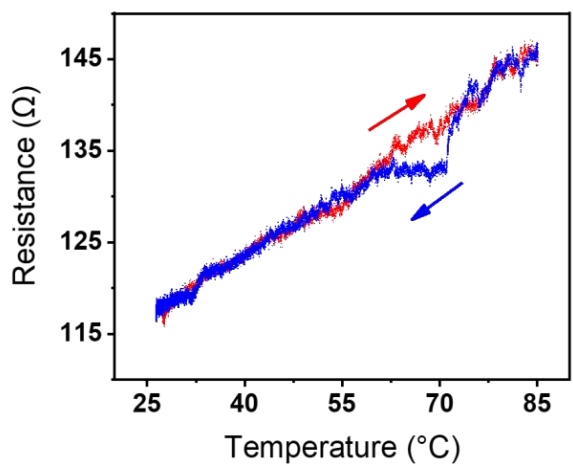

**Figure S5** Resistance vs. temperature graph of a Ag NW network device shows the bolometric response of two-terminal resistance upon heating(red) and cooling (blue).

**Comments on the origin of the alternating zero-bias signal on single Ag NW device**

Zero-bias response reported in Figure 3h-II shows successively alternating $I_{min}$ and $I_{max}$ points. Close-up SEM images of the NW shows that the nanowire has split into two segments that are very close to each other as shown in **Figure S6**. When comparing the section between the contacts to the section beyond the contacts, it can be seen that some mass has been transported between the two segments and left some undulations leading to a distinguished response. A similar response is observed due to the mass transport in a different device shown in **Figure S7**. The SPCM maps show a bipolar response at an arbitrary point of the Ag NW. After several measurements, the wire broke from the point where the photoresponse is observed. This is an interesting phenomenon that requires further investigation.

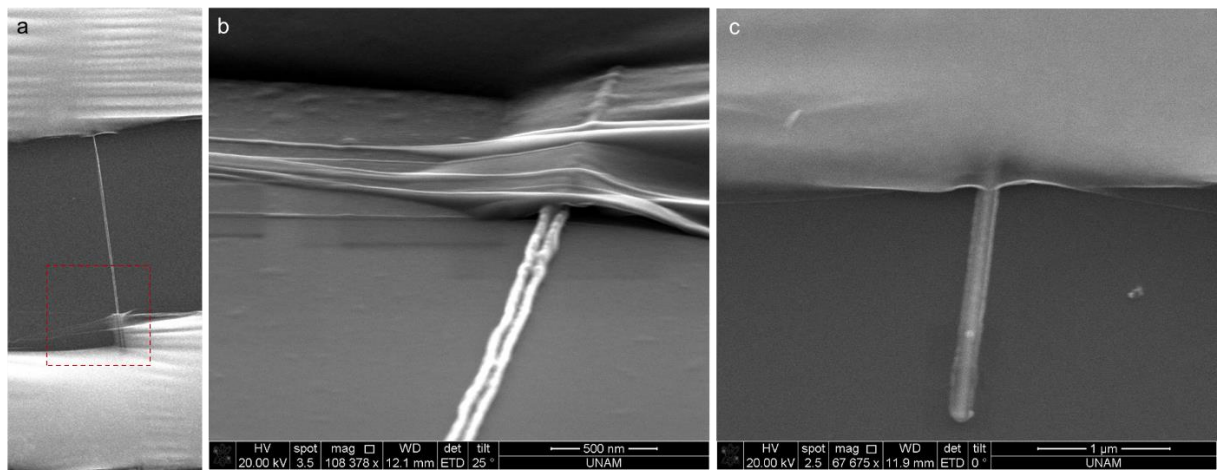

**Figure S6 a.** Mirrored SEM image of the device given in the main text. **b.** Close up view of the indium-Ag NW region after many measurements under bias. **c.** Tail of the Ag NW beyond the indium.

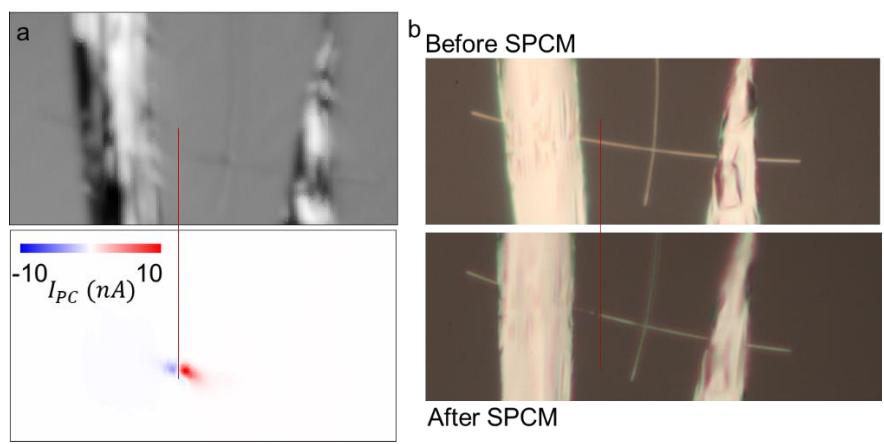

**Figure S7 a.** SPCM measurements on a single wire device. Red line is drawn to guide the eye for the spatial position of the photoresponse. **b.** Before and after optical microscope images of the device. Location of the photoresponse is marked with the red line on both images.

## Synthesis of silver nanoparticles (Ag NPs)

Ag NPs were synthesized according to a published recipe in the literature[1] (Ref. 40 in the Main Text), with some modifications. Silver nitrate (AgNO$_3$, ≥99.0%), 1-octadecene (ODE, 90%), oleylamine (OLA, 70%), ethanol (absolute), toluene (≥99.5%) and n-hexane (≥97.0%) from Sigma Aldrich are used in the synthesis process. In a typical synthesis, 30 mL of ODE, 15 mL of OLA and 600 mg of AgNO$_3$ were mixed in a 100 mL three-neck flask. The solution was kept under vacuum for 2 h at room temperature to remove the oxygen and other volatile species. Then, the flask was flushed with argon gas and the temperature was raised to 180 °C with heating rate of 7 °C/min. As the temperature was increasing, the color of the solution turned into light orange, green and finally black, which is an indicative of Ag NPs formation. The flask stayed at this temperature to have a complete nucleation and growth process, and then the temperature was decreased to 150 °C to focus the size of the Ag NPs through Ostwald ripening. After an extra 2 h, the heating mantle was removed, and the solution was diluted with hexane at room temperature. The collected solution was centrifuged at 5000 rpm for 6 min to remove the unstable or large NPs. Ag NPs were separated from the solution and unwanted species by addition of extra ethanol and centrifugation at 5000 rpm for 10 min. Finally, the precipitate was redispersed in toluene and kept in refrigerator for further use.